\documentclass[a4paper]{article}

\usepackage[margin=1in]{geometry} 

\usepackage{amsmath}
\usepackage{amsthm}
\usepackage{amssymb}
\usepackage{esint}

\usepackage[utf8]{inputenc}
\usepackage[hidelinks]{hyperref}
\hypersetup{
	unicode,
}

\usepackage[sort&compress,numbers,square]{natbib}
\theoremstyle{plain}

\theoremstyle{definition}

\def\@#1{{\mathbf{#1}}}

\def\Xint#1{\mathchoice
{\XXint\displaystyle\textstyle{#1}}%
{\XXint\textstyle\scriptstyle{#1}}%
{\XXint\scriptstyle\scriptscriptstyle{#1}}%
{\XXint\scriptscriptstyle\scriptscriptstyle{#1}}%
\!\int}
\def\XXint#1#2#3{{\setbox0=\hbox{$#1{#2#3}{\int}$ }
\vcenter{\hbox{$#2#3$ }}\kern-.575\wd0}}

\def\dashint{\Xint-}

\DeclareMathOperator{\sech}{sech}

\usepackage{graphicx, color}
\graphicspath{{fig/}}

\usepackage{comment}
\usepackage[dvipsnames,svgnames]{xcolor}

\def\gl{\mathrel{\mathpalette\overl@ss>}}

\def\sech{\mathop{\rm sech}\nolimits}

\def\Real{\mathbb{R}}

\def\Im{\mathop{\rm Im}\nolimits}
\def\Res{\mathop{\rm Res}\limits}

\def\sgn{\mathop{\rm sgn}\nolimits}

\def\@#1{{\mathbf{#1}}}
\def\_#1{{\mathsf{#1}}}
\let\~=\tilde
\let\==\bar
\let\^=\hat
\let\<=\langle
\let\>=\rangle

\def\be{\begin{equation}}
	\def\ee{\end{equation}}
\def\bse{\begin{subequations}}
	\def\ese{\end{subequations}}
\usepackage{algorithm, algpseudocode} 
\usepackage{mathrsfs} 

\usepackage{lipsum}
\numberwithin{equation}{section}
\title{Perturbation theory for kinks of the defocusing \\modified Korteweg-de Vries equation}
\author{Nicholas J. Ossi$^{1,\dagger}$, Barbara Prinari$^{1,2}$, Theodoros P. Horikis$^2$, Dimitrios J. Frantzeskakis$^3$}
\date{
    \small{$^1$ Department of Mathematics, State University of New York, Buffalo, NY 14260, USA\\
    $^2$ Department of Mathematics, University of Ioannina, Ioannina 45110, Greece\\
    $^3$ Department of Physics, National and Kapodistrian
University of Athens, Athens 15784, Greece}\\
    $^\dagger$ \href{mailto:nossi@buffalo.edu}{nossi@buffalo.edu}
}
\begin{document}
	\maketitle
	\begin{abstract}
In this work we develop an integrable perturbation theory for the defocusing modified Korteweg-de Vries kink solution based on the squared eigenfunction expansion associated with the underlying Zakharov–Shabat scattering problem. We derive the completeness relation for the squared eigenfunctions appropriate to the kink background, establish the adjoint structure needed to handle perturbations of both the continuous and discrete spectral components, and obtain explicit evolution equations for the perturbed kink parameters at leading order. The study of the first order correction shows that perturbations generically produce a radiative shelf in front of the kink. We also apply our results to certain physically relevant perturbations and show that the predictions are consistent with direct numerical simulations.
	\end{abstract}

	
\section{Introduction}

The modified Korteweg–de Vries (mKdV) equation:
\begin{equation}
    \label{e:mkdv}
q_t+6\, \sigma q^2q_x+q_{xxx}=0, \qquad \sigma=\pm 1, 
\end{equation}
where $q=q(x,t)$ is a real function of $x,t\in \Real$ and the sign of the nonlinear term distinguishes between the focusing ($+$) and defocusing ($-$) dispersion regimes, is one of the prototypical integrable nonlinear evolution equations in mathematical physics. 

Its connection to the celebrated  Korteweg-de Vries (KdV)  equation has been known since the pioneering work by Miura, who derived an explicit transformation that links solutions of mKdV to solutions of KdV \cite{Miura68}. The mKdV equation is also related to the Harry-Dym equation by a hodograph transformation \cite{Ishimori82,Kawamoto85}. In \cite{Wadati73}, Wadati solved the Cauchy problem for the focusing mKdV equation with rapidly decaying initial conditions by means of the Inverse Scattering Transform (IST), and derived its multi-soliton solutions.  In this landmark paper, Wadati 
also laid down the scattering framework to handle the non-vanishing boundary conditions that mathematically define the kink and multi-kink solutions. Explicit B\"acklund transformations specifically tailored to the mKdV equation hierarchy were constructed in \cite{Steudel74} to describe multi-kink interactions, and in \cite{Satsuma76} Satsuma extended Hirota's direct bilinear method to handle non-vanishing background steps, providing a clean method to calculate multi-kink solutions without relying on full inverse scattering equations. Explicit solutions of the mKdV equation were also obtained in terms of matrix triplets, by using the IST and solving the Marchenko equation of the inverse problem by separation of variables \cite{Demontis11}.
In \cite{Germain16} the full asymptotic stability of single solitary wave solutions for the mKdV equation under small, localized, and smooth perturbations in the energy space was established, and \cite{Chen21} proved the complete soliton resolution conjecture for the focusing mKdV equation under generic initial data in weighted Sobolev spaces, without requiring a small-norm assumption. 
Some recent works \cite{ZY20,XZF23,ZXF25} developed the IST for the focusing and defocusing mKdV equations over a constant background, and studied the long-time asymtptotics of the associated Cauchy problem, as well as the interaction of dark solitons. We also mention the more recently categorized traveling and rotating loop solitons of mKdV arising from planar geometric curve flows in \cite{Anco26}.
Among the exact solutions admitted by the defocusing mKdV equation, the kink solution occupies a distinguished position. As a topological soliton interpolating between two distinct constant asymptotic states, the kink is structurally stable and physically significant, arising naturally whenever the underlying dynamics support a bistable or symmetry-broken configuration. Understanding the behavior of the mKdV kink under perturbation — how it deforms, what radiation it emits, and whether its essential character survives — is therefore a question of both theoretical depth and practical consequence.

Since its emergence as a prototypical model of nonlinear wave phenomena, the mKdV equation — and in particular its defocusing variant — has found remarkable applicability across a broad range of physical settings. 
Indeed, the mKdV equation emerges whenever the leading quadratic nonlinearity of the KdV equation is suppressed, either by symmetry or by a near-cancellation of nonlinear effects, so that cubic nonlinearity becomes the dominant mechanism governing the dynamics. This regime is encountered in a broad spectrum of physical systems, including
acoustic waves and phonons in certain anharmonic lattices \cite{Zabusky67,Ono92}, Alfv\'en waves in cold collision-free plasmas \cite{Kakutani69,Khater98}, the dynamics of thin elastic rods \cite{Matsutani91}, the meandering of ocean currents \cite{Ralph94}, models of traffic flow dynamics \cite{Komatsu95,Ge05}, the intrinsic geometry of hyperbolic surfaces \cite{Schief95}, slag–metallic bath interfaces in metallurgical processes \cite{Agop98}, and signal propagation in Schottky barrier transmission lines \cite{Ziegler01}. This breadth of physical relevance makes the mKdV equation a compelling object of study far beyond its purely mathematical interest.

In this work, we consider the defocusing mKdV equation \eqref{e:mkdv} with $\sigma=-1$ and
subject to the nonzero boundary conditions:
\begin{equation}
\label{e:bc}
    q(x,t)\rightarrow \pm q_{0},\qquad \text{as } x\rightarrow\pm\infty,
\end{equation}
for some constant $q_0>0$. The exact kink solution of \eqref{e:mkdv}, representing a continuous transition between two constant states, is given by
\begin{equation}
    q(x,t)=q_{0}\tanh\big[q_{0}(x+2q_{0}^2t-x_{0})\big].
\end{equation}
The kink travels to the left with constant velocity fixed by the background amplitude. Its center is determined by the additional arbitrary real parameter $x_{0}$. Note that if $q(x,t)$ is a solution of \eqref{e:mkdv} so is $-q(x,t)$, so we consider $q\rightarrow\pm q_{0}$ rather than $q\rightarrow\mp q_{0}$ without loss of generality. 

The goal of this paper is to develop an integrable perturbation theory for the defocusing mKdV kink solution to predict the slow-time evolution of the kink velocity and center, as well as the radiation shelf emerging on its side when small perturbations (e.g., diffusion, linear/nonlinear loss, etc.) are considered. The development of systematic perturbation theories for integrable equations has a long history, with the IST providing the natural backdrop against which corrections can be organized. A particularly elegant and powerful approach to this problem exploits the so-called squared eigenfunction basis, which arises from the linearization of the underlying Lax pair structure.
However, the literature on perturbation theory specifically for the mKdV equation — as opposed to KdV, nonlinear Schr\"odinger (NLS), or sine-Gordon — is surprisingly sparse, and even more so for the mKdV dark solitons (i.e., solitons on a non-zero background) and for the kink. Although the broader treatments of  Karpman \& Maslov \cite{KM77} and Kaup \& Newell \cite{KN78} cover the focusing mKdV bright (i.e., rapidly decaying) soliton implicitly, since the mKdV belongs to the AKNS hierarchy, they do not single it out for detailed treatment. The comprehensive review \cite{KM89} by Kivshar and Malomed, surveys perturbation results for the KdV, NLS, sine-Gordon, and Landau–Lifshitz equations, but not mKdV specifically. The work by Yang
\cite{Yang03} identified stable embedded solitons in the generalized third-order 
 NLS equation. Upon reducing this equation to a perturbed complex mKdV equation, Yang constructed a soliton perturbation theory showing that a continuous family of sech-shaped embedded solitons (on a zero background) exists and exhibits nonlinear stability, thereby establishing that embedded solitons can be just as robust as conventional solitons.
Another relevant work is \cite{Lin13}, which studies what happens to an mKdV bright soliton when a small non-Hamiltonian perturbation — specifically, a multiplication by a small spatially varying potential — is added to the equation. The main findings are two-fold. First, the soliton is orbitally stable under this perturbation: even though the perturbation is not Hamiltonian (so standard energy-based methods do not directly apply), the solution remains close to the soliton manifold over a long time scale, with the deviation growing only slowly. On a somewhat shorter but still dynamically relevant time scale, the position and amplitude of the soliton evolve according to a simple and explicit system of ordinary differential equations, allowing one to predict the soliton's trajectory. A notable consequence is that the perturbation can produce an order-one shift in the soliton's position over this time scale, even though it is small. 
 he only study that appears to deal explicitly with perturbations of mKdV kinks is Ref.~\cite{Chen18}, which employs geometric singular perturbation theory to prove the existence of kink and periodic traveling-wave solutions in a perturbed defocusing mKdV model. By analyzing the perturbation of the Hamiltonian vector field with an elliptic Hamiltonian of degree four, a two-saddle cycle is exhibited, the wave speed $c_o(h)$ is shown to be decreasing on $h\in[-3/4,0]$ by analysis of the ratio of Abelian integrals, and the relationship between wave speed and wavelength of the resulting traveling waves is determined. The key finding is that the kink solution persists under small dissipative or dispersive perturbations, with its speed modified in a computable way. 

Among the various perturbative approaches (e.g., multi-scale analysis, perturbations of conserved quantities, etc), IST-based eigenfunction-expansion techniques stand out for their elegance: squared eigenfunctions diagonalize the linearized flow, provide a resolution of the identity for the relevant function spaces, and allow corrections to scattering data to be expressed in closed form. Of course, the crux of the matter is whether the completeness of the squared eigenfunctions can be established.
For a detailed treatment of the squared-eigenfunction expansion method, see Chapter 4 of \cite{Yang} and the references therein. This approach has been successfully applied over the decades to study perturbations of bright solitons of the focusing NLS equation, the KdV equation and related hierarchies, 
but its adaptation to the case of dark solitons (or, more generally, coherent structures on a nontrivial background) presented some serious challenges.
Recently, in \cite{OPY26} the method was applied to dark solitons of the defocusing NLS equation 
(see also \cite{KonotopPRE1994,Chen98,Chen99,Ao05,Ao06,Ao07} for earlier results). Its implementation for the defocusing mKdV kink — whose associated completeness relation has its own distinctive structure — merits a careful and self-contained treatment. Specifically, we derive the completeness relation for the squared eigenfunctions appropriate to the kink background, establish the adjoint structure needed to handle perturbations of both the continuous and discrete spectral components, and obtain explicit evolution equations for the perturbed kink parameters at leading order. The completeness relation is obtained by adapting the result of \cite{OPY26} for the Zakharov-Shabat scattering problem shared by the NLS and mKdV equations. 
This is
possible since the completeness result only involves time-independent spectral data, and follows the approach used in \cite{Y2000,ABC2022} for the case of zero boundary conditions. 
On the other hand, unlike NLS, mKdV is a real nonlinear partial differential equation (PDE), and one needs to account for the additional symmetry in the spectral variable ensuring that $q(x,t)\in \Real$ for all $x,t\in \Real$, which in turn modifies the completeness relation. 

The paper is organized as follows. Section 2 reviews the IST for the defocusing mKdV equation and its kink solution. In Section 3 we introduce the squared eigenfunction basis, and derive the completeness and orthogonality relations. Section 4 develops the perturbation theory and derives the modulation equations for the kink. In Section~5 we analyze  the first order correction and show that perturbations generically produce a prominent shelf on the left side (i.e., in front) of the kink.  Section 6 presents illustrative applications, and Section 7 contains concluding remarks.

\section{Overview of the IST}

The IST for the defocusing mKdV equation with nonvanishing boundary conditions is described in detail in \cite{ZY20}. Here, we summarize the ingredients that will be important for the present study while omitting most details. Note that the process is nearly identical to the better-known IST for the defocusing nonlinear Schr\"odinger equation with nonzero boundary conditions (see \cite{P2023} for a review), only with one additional symmetry in the scattering data that ensures the reality of the potential. The Lax pair of \eqref{e:mkdv} is given by
\bse
\begin{align}
    &\varphi_{x}=U\varphi,\qquad U=-ik\sigma_{3}+q\sigma_{1},\quad \sigma_{1}=\begin{bmatrix}
        0&1\\1&0
    \end{bmatrix},\quad\sigma_{3}=\begin{bmatrix}
        1&0\\0&-1
    \end{bmatrix},\\
    &\varphi_{t}=V\varphi,\qquad V=4k^2U+2ik\sigma_{3}(q_{x}\sigma_{1}-q^{2}I)+2q^{3}\sigma_{1}-q_{xx}\sigma_{1}.
\end{align}
\ese
In \cite{ZY20}, the IST is laid out for \eqref{e:mkdv} subject to the boundary conditions $q\rightarrow q_{\pm}$ as $x\rightarrow\pm\infty$ with $|q_{\pm}|=q_{0}$, where there are two distinct possibilities: $q_{\pm}=q_{0}$ and $q_{\pm}=\pm q_{0}$. In what follows, we adapt the presentation of \cite{ZY20} while assuming the latter choice for the boundary values, since we are interested in the kink solution. As $x\rightarrow\pm\infty$, given the prescribed boundary conditions \eqref{e:bc} we have
\begin{equation}
    U\sim \begin{bmatrix}
        -ik&\pm q_{0}\\\pm q_{0}&ik
    \end{bmatrix},\qquad V\sim\begin{bmatrix}
        -ik&\pm q_{0}\\\pm q_{0}&ik
    \end{bmatrix}(4k^2+2q_{0}^{2}).
\end{equation}
Thus, the asymptotic eigenvalues of $U$ are $\pm i\lambda$ with $\lambda^{2}=k^2-q_{0}^{2}$, and the asymptotic eigenvalues of $V$ are $\pm i\lambda(4k^2+2q_{0}^2)$, and both have branch points at $k=\pm q_0$. To avoid dealing with branching, one can define the uniformization variable $z$ such that
\begin{equation}
    z=k+\lambda,\qquad k=\frac{1}{2}(z+q_{0}^{2}z^{-1}),\qquad\lambda=\frac{1}{2}(z-q_{0}^{2}z^{-1}),
\end{equation}
which maps both sheets of the Riemann surface on which $k$ is defined into the complex $z$-plane (see \cite{OPY26} for details). 
Then, the Jost eigenfunctions are defined according to the boundary conditions
\bse
\begin{align}
    &\Phi(x,t,z)=\begin{bmatrix}
        \phi(x,t,z)&\bar\phi(x,t,z)
    \end{bmatrix}\sim E_{-}(z)e^{-i\Omega(x,t,z)\sigma_{3}},\quad\text{as}\;x\rightarrow-\infty,\\
    &\Psi(x,t,z)=\begin{bmatrix}
        \bar\psi(x,t,z)&\psi(x,t,z)
    \end{bmatrix}\sim E_{+}(z)e^{-i\Omega(x,t,z)\sigma_{3}},\quad\text{as}\;x\rightarrow+\infty,
\end{align}
\ese
where
\begin{equation}
    E_{\pm}(z)=\begin{bmatrix}
        1&\mp iq_{0}/z\\
        \pm iq_{0}/z&1\end{bmatrix},\qquad\Omega(x,t,z)=\lambda\big[x+(4k^2+2q_{0}^2)t\big].
\end{equation}
Note that
\begin{equation}
    \det\Phi=\det\Psi=1-q_{0}^2z^{-2}=:\gamma(z).
\end{equation}
It can be proven that if the potential $q(x,t)$ decays to the background $q_\pm$ sufficiently rapidly as $x\to \pm \infty$, the eigenfunctions $\phi$ and $\psi$ can be analytically extended into the upper-half $z$ plane, while $\bar\phi$ and $\bar\psi$ are analytic in the lower-half plane. The Jost eigenfunctions satisfy the following symmetries corresponding to the involutions $z\mapsto z^{*}$, $z\mapsto q_{0}^{2}/z$, and $z\mapsto-z$ respectively:
\bse
\begin{align}
\label{e:sym1}
    \Phi(x,t,z)=\sigma_{1}\Phi^{*}(x,t,z^*)\sigma_{1},\qquad&\Psi(x,t,z)=\sigma_{1}\Psi^{*}(x,t,z^{*})\sigma_{1},\\
\label{e:sym2}
    \Phi(x,t,z)=\frac{iq_{0}}{z}\Phi(x,t,q_{0}^{2}/z)\sigma_{3}\sigma_{1},\qquad&\Psi(x,t,z)=-\frac{iq_{0}}{z}\Psi(x,t,q_{0}^{2}/z)\sigma_{3}\sigma_{1},\\
\label{e:sym3}
        \Phi(x,t,z)=\sigma_{1}\Phi(x,t,-z)\sigma_{1},\qquad&\Psi(x,t,z)=\sigma_{1}\Psi(x,t,-z)\sigma_{1}.
\end{align}
\ese
Note that the symmetry \eqref{e:sym3} is unique to the mKdV equation and is necessary to ensure reality of the potential. Combining \eqref{e:sym2} and \eqref{e:sym3} yields the useful relationships between the components of $\phi$ and $\psi$,
\begin{equation}
\label{e:components_sym}
    \phi_{1}(x,t,z)=-\frac{iq_{0}}{z}\phi_{2}(x,t,-q_{0}^{2}/z),\qquad  \psi_{1}(x,t,z)=-\frac{iq_{0}}{z}\psi_{2}(x,t,-q_{0}^{2}/z).
\end{equation}
The Jost eigenfunctions can be related via a scattering matrix for $z\in\mathbb{R}\setminus
\{\pm q_{0}\}$:
\begin{equation}
\label{e:scat}
    \Phi(x,t,z)=\Psi(x,t,z)S(z),\qquad S(z)=\begin{bmatrix}
        a(z)&\bar{b}(z)\\
        b(z)&\bar{a}(z)
    \end{bmatrix},\qquad \det S(z)=1.
\end{equation}
The so-called scattering coefficients can then be expressed as
\begin{equation}
    a(z)=\frac{1}{\gamma(z)}W(\phi,\psi),\qquad    \bar a(z)=\frac{1}{\gamma(z)}W(\bar\psi,\bar\phi), \qquad   b(z)=\frac{1}{\gamma(z)}W(\bar\psi,\phi),    \qquad\bar{b}(z)=\frac{1}{\gamma(z)}W(\bar\phi,\psi),
\end{equation}
where $W(f,g)$ denotes the Wronskian of two vector functions $f,g$. Thus, $a(z)$ can be analytically extended into the upper-half plane, while $\bar{a}(z)$ can be extended into the lower-half plane. Importantly, the scattering coefficients are generically singular at the branch points $\pm q_{0}$ on the real axis, unless the Jost eigenfunctions become linearly dependent (which is indeed the case for a pure kink solution). Moreover, the scattering coefficients satisfy the symmetries
\bse
\begin{align}
    &{a}(z)=\bar a^{*}(z^{*})=-\bar{a}(q_{0}^{2}/z)=\bar{a}(-z),\qquad\Im z\geq0,\\&{b}(z)=\bar b^{*}(z)=\bar{b}(q_{0}^{2}/z)=\bar{b}(-z),\qquad\qquad z\in\mathbb{R}.
\end{align}
\ese
The inverse scattering problem is formulated as a Riemann-Hilbert problem across the real $z$ axis. To this end, \eqref{e:scat} is written as
\begin{equation}
    \frac{\phi(x,t,z)}{a(z)}=\bar\psi(x,t,z)+\rho(z)\psi(x,t,z),\qquad\frac{\bar\phi(x,t,z)}{\bar a(z)}=\psi(x,t,z)+\bar\rho(z)\bar\psi(x,t,z),
\end{equation}
where the reflection coefficients are defined as: 
\begin{equation}
\rho(z)=\frac{b(z)}{a(z)}, \qquad \bar{\rho}(z)=\frac{\bar{b}(z)}{\bar{a}(z)}.
\end{equation}
It is known that $a(z)$ has a finite number of simple complex zeros that lie in the upper-half plane on the circle $|z|=q_{0}$. However, we are presently interested in the kink solution satisfying the boundary condition \eqref{e:bc}. This solution corresponds to the special case where $a(z)$ has only the purely imaginary zero $iq_{0}$ (and $\bar{a}(z)$ has the corresponding zero $-iq_{0}$). Note that eigenvalues off the imaginary axis correspond to dark solitons on a constant background satisfying the boundary condition $q\rightarrow q_{0}$ as $x\rightarrow\pm\infty$. Due to the symmetries of the scattering coefficients, such discrete eigenvalues appear in symmetric quartets $\{\zeta_{j},-\zeta_{j}^{*},\zeta_{j}^{*},-\zeta_{j}\}$. 
In the purely imaginary case corresponding to the kink solution considered here, this quartet collapses into the conjugate pair $\{iq_{0},-iq_{0}\}$. At the discrete eigenvalue, the Jost eigenfunctions satisfy
\begin{equation}
    \phi(x,t,iq_{0})=b_{0}\psi(x,t,iq_{0}),\qquad\bar\phi(x,t,-iq_{0})=\bar{b}_{0}\bar\psi(x,t,-iq_{0}),
\end{equation}
for some constants satisfying $b_{0}=\bar{b}_{0}^{*}=\bar{b}_{0}$. Furthermore, the residue contributions from the discrete eigenvalue are given by
\begin{equation}
    \Res_{z=iq_{0}}\frac{\phi(x,t,z)}{a(z)}=C_{0}\psi(x,t,iq_{0}),\qquad\Res_{z=-iq_{0}}\frac{\bar\phi(x,t,z)}{\bar{a}(z)}=\bar{C}_{0}\bar\psi(x,t,-iq_{0}).
\end{equation}
The norming constants are defined by $C_{0}=b_{0}/a'(iq_{0})$ and $\bar C_{0}=\bar b_{0}/\bar a'(-iq_{0})$, where the prime denotes a derivative with respect to $z$. They can be shown to satisfy the symmetries $C_{0}=\bar{C}_{0}^{*}=-\bar C_{0}$. Thus, in the pure kink (reflectionless) case the Riemann-Hilbert problem reduces to the linear system
\bse
\label{e:system}
\begin{align}
    \psi(x,t,z)e^{-i\Omega(x,t,z)}&=\begin{bmatrix}
        -iq_{0}/z\\1
    \end{bmatrix}+\frac{e^{-i\Omega(x,t,-iq_{0})}\bar\psi(x,t,-iq_{0})C_{0}^*}{z+iq_{0}},\\
    \bar\psi(x,t,z)e^{i\Omega(x,t,z)}&=\begin{bmatrix}
        1\\iq_{0}/z
    \end{bmatrix}+\frac{e^{i\Omega(x,t,iq_{0})}\psi(x,t,iq_{0})C_{0}}{z-iq_{0}}.
\end{align}
\ese
With the fact that the norming constants are purely imaginary in mind, we introduce the new parameter $x_{0}\in\mathbb{R}$ through
\begin{equation}
    C_{0}=-2iq_{0}e^{2q_{0}x_{0}}.
\end{equation}
The sign was chosen to ensure a non-singular solution. Then, we introduce the traveling coordinate
\begin{equation}
\label{e:xi}
    \xi=q_{0}(x+2q_{0}^{2}t-x_{0}),
\end{equation}
and write the solution of the above system \eqref{e:system} in the form:
\bse
\begin{align}
\psi(x,t,z)&=\begin{bmatrix}
    q_{0}^{2}/z-iq_{0}\tanh\xi\\
    z+iq_{0}\tanh\xi
\end{bmatrix}\frac{e^{i\Omega(x,t,z)}}{z+iq_{0}},\\
    \bar\psi(x,t,z)&=\begin{bmatrix}
        z-iq_{0}\tanh\xi\\
        q_{0}^{2}/z+iq_{0}\tanh\xi
    \end{bmatrix}\frac{e^{-i\Omega(x,t,z)}}{z-iq_{0}}.
\end{align}
\ese
The rest of the Jost eigenfunctions can be obtained from
\begin{equation}
    \phi(x,t,z)=a(z)\bar\psi(x,t,z),\qquad\bar\phi(x,t,z)=\bar{a}(z)\psi(x,t,z),\qquad a(z)=\frac{z-iq_{0}}{z+iq_{0}},\qquad\bar{a}(z)=\frac{z+iq_{0}}{z-iq_{0}}.
\end{equation}
Finally, the potential can be recovered through
\begin{equation}
    q(x,t)=\lim_{z\rightarrow\infty}\left( iz\psi_{1}(x,t,z)e^{-i\Omega(x,t,z)} \right),
\end{equation}
which yields the kink
\begin{equation}
    q(x,t)=q_{0}\tanh\xi.
\end{equation}

\section{Squared eigenfunctions}

In \cite{OPY26}, a completeness relation for certain quadratic combinations of Jost eigenfunctions, so-called squared eigenfunctions, was derived for the defocusing nonlinear Schr\"odinger equation with nonzero boundary conditions. Since the mKdV equation is associated with the same Zakharov-Shabat scattering problem, this result can be adapted to the present study. In particular, the following completeness result was established in \cite{OPY26}:
\begin{equation}
\label{e:completeness_nls}
    \begin{bmatrix}
        f(x)\\f^*(x)
    \end{bmatrix}=\frac{1}{2\pi}\int_{\Gamma}d\zeta\frac{1}{\gamma(\zeta)a(\zeta)^2}\eta(x,t,\zeta)\int_{-\infty}^{\infty}dy\,\chi(y,t,\zeta)^T\begin{bmatrix}
        f(y)\\f^{*}(y)
    \end{bmatrix},
\end{equation}
where $\Gamma$ is a contour in the upper-half plane from $-\infty+i0$ to $+\infty+i0$ passing above the circle of radius $q_{0}$, the superscript $^T$ denotes matrix transpose, and the squared eigenfunction $\eta$ and adjoint squared eigenfunction $\chi$ are given by
\begin{equation}
    \eta=\begin{bmatrix}
        \psi_{1}^2\\
        \psi_{2}^2
    \end{bmatrix},\qquad\chi=\begin{bmatrix}
        -\phi_{2}^2\\
        \phi_{1}^2
    \end{bmatrix},
\end{equation}
in terms of the components $\psi_j$ and $\phi_j$ for $j=1,2$ of the Jost eigenfunctions defined in Section~2. 
For the purpose of studying perturbation theory for the mKdV, we only need to be able to expand real scalar functions, and as such the completeness relation can be reduced to a scalar form. This is done in the context of the focusing mKdV equation with zero boundary conditions in \cite{Y2000} (see also \cite{ABC2022}). With this in mind, adding the two components of \eqref{e:completeness_nls} yields
\begin{equation}
    f(x)=\frac{1}{4\pi}\int_{\Gamma}d\zeta\frac{1}{\gamma(\zeta)a(\zeta)^2}\eta_{s}(x,t,\zeta)\langle\chi_{s}(t,\zeta),f\rangle,
\end{equation}
 where the standard spatial inner product is defined as $\langle g, f\rangle=\int_{-\infty}^{\infty}g(y)f(y)dy$, and the scalar squared eigenfunctions are
 \begin{equation}
 \label{e:se_scalar}
     \eta_{s}(x,t,z)=\psi_{1}(x,t,z)^2+\psi_{2}(x,t,z)^2,\qquad \chi_{s}(x,t,z)=\phi_{1}(x,t,z)^2-\phi_{2}(x,t,z)^2.
 \end{equation}
 Equivalently, we have the closure/completeness relation
 \begin{equation}
     \delta(x-y)=\frac{1}{4\pi}\int_{\Gamma}\frac{\eta_{s}(x,t,\zeta)\chi_{s}(y,t,\zeta)}{\gamma(\zeta)a(\zeta)^2}d\zeta.
 \end{equation}
 We now specialize the closure relation to a pure kink solution. In this case, it is convenient to express everything in terms of the traveling variable $\xi$ defined in \eqref{e:xi}. In the reference frame of the kink, we can introduce the time-independent scalar squared eigenfunctions by
\bse
\label{e:se_mod}
\begin{align}
    Z(\xi,z)&=\eta_{s}(x,t,z)e^{-2i\lambda(4k^2t+x_{0})}(z+iq_{0})^2,\\
    Y(\xi,z)&=\chi_{s}(x,t,z)e^{2i\lambda(4k^2t+x_{0})}(z+iq_{0})^2,
\end{align}
\ese
which satisfy the closure relation
\begin{equation}
    q_{0}\delta(\xi-\xi')=\frac{1}{4\pi}\int_{\Gamma}\frac{Z(\xi,\zeta)Y(\xi',\zeta)}{\gamma(\zeta)(\zeta^2+q_{0}^2)^2}d\zeta.
\end{equation}
Straightforward calculations show that the modified squared eigenfunctions have the explicit expressions 
\bse
\begin{align}
    Z(\xi,z)&=\left[(q_{0}^{2}/z-iq_{0}\tanh\xi)^2+(z+iq_{0}\tanh\xi)^2\right]e^{2i\lambda q_{0}^{-1}\xi},\\
    Y(\xi,z)&=\left[(z-iq_{0}\tanh\xi)^2-(q_{0}^{2}/z+iq_{0}\tanh\xi)^2\right]e^{-2i\lambda q_{0}^{-1}\xi}.
\end{align}
\ese
We now explicitly extract the residue from the discrete eigenvalue $iq_{0}$. Note that $Y(\xi,iq_{0})=0$, so in this formulation the discrete eigenvalue is a simple pole of the integrand, rather than a double pole. The residue is given by
\begin{equation}
    \Res_{z=iq_{0}}\frac{Z(\xi,z)Y(\xi',z)}{\gamma(z)(z^2+q_{0}^2)^2}=-\frac{1}{8q_{0}^2}Z_{0}(\xi)Y_{0}(\xi'),
\end{equation}
where we define the discrete eigenfunctions as
\begin{equation}
    Z_{0}(\xi):=Z(\xi,iq_{0}),\qquad Y_{0}(\xi):=\partial_{z}Y(\xi,z)\big\vert_{z=iq_{0}},
\end{equation}
with the explicit expressions:
\begin{align}
    Z_{0}(\xi)=-2q_{0}^2\sech^2\xi,\qquad
    Y_{0}(\xi)=4iq_{0}e^{\xi}\sech\xi.
\end{align}
Thus, the closure relation can be written as
\begin{equation}
\label{e:completeness_kink}
    q_{0}\delta(\xi-\xi')=\frac{1}{4\pi}\int_{C}\frac{Z(\xi,\zeta)Y(\xi',\zeta)}{\gamma(\zeta)(\zeta^2+q_{0}^2)^2}d\zeta+\frac{i}{16q_{0}^2}Z_{0}(\xi)Y_{0}(\xi'),
\end{equation}
where $C$ is a contour along the real axis that is indented in the upper-half plane to avoid the poles at $\pm q_{0}$ that are present due to the fact that $\gamma(\pm q_0)=0$. Note that in \cite{OPY26}, additional half residues from $\pm q_{0}$ were explicitly extracted at this point. We will later demonstrate that this is not necessary for the present study. Unlike the NLS case, when computing the first order term in the perturbation expansion which describes the radiation shelf that develops in front of the perturbed kink, the singularities become removable and the contour can be deformed to the real axis.

\section{Perturbation theory}

We now consider the defocusing mKdV subject to a small perturbation:
\begin{equation}
\label{e:mkdv_pert}
    q_{t}(x,t)-6q(x,t)^2q_{x}(x,t)+q_{xxx}(x,t)=\varepsilon F[q(x,t)],\qquad0<\varepsilon\ll1.
\end{equation}
We aim to describe the dynamics of a single kink under the influence of the perturbation. To this end, we introduce an ansatz of the form
\begin{equation}
\label{e:ansatz}
    q(x,t)=
    u(\xi)+\varepsilon\tilde{q}(\xi,t) +\mathcal{O}(\varepsilon^2),
\end{equation}
where $u(\xi)=q_{0}\tanh\xi$ is the exact kink solution and $\tilde{q}(\xi,t)$ is the first-order correction. Additionally, we allow the kink parameters to evolve on a slow time scale $T=\varepsilon t$,
\begin{equation}
    q_{0}=q_{0}(T),\qquad x_{0}=x_{0}(T),\qquad\xi=q_{0}(T)\left(x+2\int_{0}^{t}q_{0}(\varepsilon s)^{2}ds-x_{0}(T)\right).
\end{equation}
Throughout, we often omit writing the dependence on $T$ for convenience. It is worth noting at this point that the slow evolution of the background amplitude $q_{0}(T)$ can be determined \textit{a priori} by directly taking a limit of \eqref{e:mkdv_pert} as $\xi\rightarrow\infty$, which yields
\begin{equation}
\label{e:q0T}
    q_{0T}=F[q_{0}],
\end{equation}
where subscript $_T$ denotes the derivative with respect to the slow time $T$.
This result is equivalently produced using an orthogonality condition as a part of the method we now describe. Inserting the ansatz \eqref{e:ansatz} into \eqref{e:mkdv_pert} gives a linearized problem at $\mathcal{O}(\varepsilon)$ of the form
\begin{equation}
    \mathcal{L}\tilde q(\xi,t)=W(\xi),
\end{equation}
where the right-hand side is given explicitly by
\begin{equation}
    W(\xi)=F[u(\xi)]-\partial_{T}u(\xi)=F[u(\xi)]-q_{0T}(\tanh\xi+\xi\sech^{2}\xi)+q_{0}^2x_{0T}\sech^{2}\xi.
\end{equation}
The linearization operator $\mathcal{L}$ and its formal adjoint $\mathcal{L}^{A}$ are given by
\begin{equation}
    \mathcal{L}=\partial_{t}-6\partial_{x}\big[u(\xi)^2\cdot\big]+\partial_{x}^{3},\qquad \mathcal{L}^{A}=-\partial_{t}+6u(\xi)^2\partial_x-\partial_{x}^{3}.
\end{equation}
One can verify directly that the scalar squared eigenfunctions as defined in \eqref{e:se_scalar} satisfy
\begin{equation}
\mathcal{L}\eta_{s}(x,t,z)=0,\qquad\mathcal{L}^{A}\chi_{s}(x,t,z)=0.
\end{equation}
In the traveling reference frame of the kink (while allowing for additional explicit dependence on $t$), we have $\mathcal{L}=\partial_{t}+L$ and $\mathcal{L}^{A}=-\partial_{t}+L^{A}$ with 
\begin{equation}
    L=2q_{0}^{3}\partial_{\xi}-6q_{0}\partial_{\xi}\big[u(\xi)^2\cdot\big]+q_{0}^{3}\partial_{\xi}^{3},\qquad {L}^{A}=-2q_{0}^{3}\partial_{\xi}+6q_{0}u(\xi)^2\partial_\xi-q_{0}^{3}\partial_{\xi}^{3},
\end{equation}
and the modified squared eigenfunctions as in \eqref{e:se_mod} are shown to satisfy
\begin{equation}
    LZ(\xi,z)=-8i\lambda k^2Z(\xi,z), \qquad L^AY(\xi,z)=-8i\lambda k^2Y(\xi,z).
\end{equation}
Furthermore, the discrete modes satisfy $LZ_{0}(\xi)=0$ and $L^{A}Y_{0}(\xi)=0$. As such, the modified squared eigenfunctions provide a natural basis in which to expand the linearized problem $(\partial_{t}+L)\tilde{q}(\xi,t)=W(\xi)$. Using the completeness relation \eqref{e:completeness_kink}, we expand the first-order correction as
\begin{equation}
    \tilde{q}(\xi,t)=\frac{1}{4\pi}\int_{C}\frac{g(t,\zeta)}{\gamma(\zeta)(\zeta^2+q_{0}^2)^2}Z(\xi,\zeta)d\zeta+\frac{i}{16q_{0}^{2}}g_{0}(t)Z_{0}(\xi).
\end{equation}
Substituting this expression into the linearized problem, similarly expanding $W(\xi)$, and equating coefficients gives
\begin{equation}
    \partial_{t}g(t,\zeta)-8i\lambda k^2g(t,\zeta)=\langle Y(\zeta),W\rangle,\qquad\partial_{t}g_{0}(t)=\langle Y_{0},W\rangle.
\end{equation}
The second equation above implies that $g_{0}(t)=\langle Y_{0},W\rangle t$, which produces secular growth in the first-order correction. To suppress this, the following orthogonality condition must be satisfied:
\begin{equation}
\label{e:orth}
    \langle Y_{0},W\rangle=0.
\end{equation}
The inner product above is generically divergent and, as such, this orthogonality condition actually provides two constraints on the kink parameters. The removal of divergence from the inner product provides the evolution of the background amplitude $q_{0T}$ (which necessarily must agree with \eqref{e:q0T}), while enforcing that the remaining convergent terms vanish gives the evolution of the center $x_{0T}$. 
Specifically, evaluating the convergent integrals, the condition \eqref{e:orth} reads
\begin{equation}
    \int_{-\infty}^{\infty}e^{\xi}\sech\xi\Big\{F[u(\xi)]-q_{0T}\tanh\xi\Big\}d\xi-q_{0T}+2q_{0}^{2}x_{0T}=0,
\end{equation} 
showing that evolution of $q_{0}$ must be chosen to remove the divergence, while the evolution of $x_{0}$ is determined from the convergent terms. That said, it is not appropriate to split this condition into two separate evolution equations until the form of the perturbation $F$ has been specified, since the integral above could include both divergent and convergent parts.
Next, solving the ordinary differential equation for $g(t,\zeta)$ with zero initial condition gives an integral representation of the first-order correction,
\begin{equation}
\label{e:foc}
    \tilde{q}(\xi,t)=\frac{i}{64\pi}\int_{C}\frac{\zeta\langle Y(\zeta),W\rangle}{k^2(\zeta^2+q_{0}^2)^2}\frac{1-e^{8i\lambda k^2t}}{\lambda^2}Z(\xi,\zeta)d\zeta,
\end{equation}
where we have also used $\gamma(z)=2\lambda(z)/z$. In the following section, we will show that this first-order correction predicts the development of a propagating shelf on one side of the kink.

\section{First-order correction}

We now study the first-order correction integral \eqref{e:foc} in the large $|\xi|$ asymptotic limit. First, for $\xi\rightarrow+\infty$ we have $Z(\xi,\zeta)\sim4\lambda(\lambda+iq_{0})e^{2i\lambda q_{0}^{-1}\xi}$, so that 
\begin{equation}
\label{e:foc_2}
    \tilde{q}^{+}\sim\frac{i}{16\pi}\int_{-\infty}^{\infty}\frac{\zeta\langle Y(\zeta),W\rangle}{k^2(\zeta^2+q_{0}^2)^2}\frac{1-e^{8i\lambda k^2t}}{\lambda}(\lambda+iq_{0})e^{2i\lambda q_{0}^{-1}\xi}d\zeta,
\end{equation}
where the last exponential factor is rapidly oscillatory in the large $|\xi|$ limit. Note that the integration over the indented contour $C$ has been replaced with integration over the real axis, since the singularities at $\pm q_{0}$ are removable. However, it will now be useful to split the integral into two parts, each possessing poles at $\pm q_{0}$ and interpreted in the Cauchy principal value sense. In particular, we write
\begin{equation}
    \tilde{q}^{+}\sim\dashint_{-\infty}^{\infty}\mathcal{F}(\zeta)\frac{e^{2i\lambda q_{0}^{-1}\xi}}{\lambda}d\zeta-\dashint_{-\infty}^{\infty}\mathcal{F}(\zeta)\frac{e^{8i\lambda k^2t+2i\lambda q_{0}^{-1}\xi}}{\lambda}d\zeta,\qquad    \mathcal{F}(z)=\frac{iz(\lambda+iq_{0})}{16\pi k^2(z^2+q_{0}^2)^2}\langle Y(z),W\rangle,
\end{equation}
where $\dashint$ denotes the principal value integral.
To asymptotically approximate these rapidly oscillatory principal value integrals, we use the fact that the dominant contributions to the integrals come from the neighborhoods of the poles $\pm q_{0}$, and then employ the identity
\begin{equation}
    \dashint_{-\infty}^{\infty}\frac{e^{ip\vartheta}}{p}dp=\pi i\sgn\vartheta.
\end{equation}
Noting that $\lambda\approx z\mp q_{0}$ and $k\approx\pm q_{0}$ near $z=\pm q_{0}$, this gives
\begin{equation}
    \tilde{q}^{+}\sim\pi i\sum_{\pm}\mathcal{F}(\pm q_{0})\Big\{\sgn\xi-\sgn[q_{0}^{-1}\xi+4q_{0}^{2}t]\Big\}=0,
\end{equation}
since $\xi$ and $t$ are both positive in this limit. Thus, there is no shelf on the right side of the kink. This is expected due to the fact that the mKdV equation only allows unidirectional propagation. Now, consider the opposite asymptotic limit $\xi\rightarrow-\infty$ in which case $Z(\xi,\zeta)\sim4\lambda(\lambda-iq_{0})e^{2i\lambda q_{0}^{-1}\xi}$ and
\begin{equation}
    \tilde{q}^{-}\sim\dashint_{-\infty}^{\infty}\tilde{\mathcal{F}}(\zeta)\frac{e^{2i\lambda q_{0}^{-1}\xi}}{\lambda}d\zeta-\dashint_{-\infty}^{\infty}\tilde{\mathcal{F}}(\zeta)\frac{e^{8i\lambda k^2t+2i\lambda q_{0}^{-1}\xi}}{\lambda}d\zeta,\qquad    \tilde{\mathcal{F}}(z)=\frac{iz(\lambda-iq_{0})}{16\pi k^2(z^2+q_{0}^2)^2}\langle Y(z),W\rangle.
\end{equation}
Similarly to the above, we find
\begin{equation}
\label{e:q_min}
    \tilde{q}^{-}\sim\pi i\sum_{\pm}\tilde{\mathcal{F}}(\pm q_{0})\Big\{\sgn\xi-\sgn[q_{0}^{-1}\xi+4q_{0}^{2}t]\Big\}.
\end{equation}
In this case $\xi$ is negative, and in the regime $q_{0}^{-1}\xi\ll-4q_{0}^2t$ we can conclude that $\tilde{q}^{-}\sim0$. On the other hand, in the region $-4q_{0}^2t\ll q_{0}^{-1}\xi\ll-1$ the first-order correction is asymptotically non-negligible. This corresponds to a prominent shelf on the left side (i.e., in front) of the kink. In terms of the original spatial variable, 
the shelf exists in the region
\begin{equation}
    -6q_{0}^2t\ll x-x_{0}\ll -2q_{0}^{2}t-1.
\end{equation}
Thus, the edge of the shelf propagates to the left with a velocity of $-6q_{0}^2$. Note that the shelf speed is strictly greater in magnitude than that of the kink ($-2q_{0}^2$), so the kink cannot overtake the shelf. We can obtain a prediction for the height of the shelf by evaluating \eqref{e:q_min}. The adjoint squared eigenfunction evaluated at the branch points is found to be
\begin{equation}
    Y(\xi,\pm q_{0})=\mp4iq_{0}^{2}\tanh\xi=\mp4iq_{0}u(\xi),
\end{equation}
which then gives
\begin{equation}
\label{e:alpha}
    \tilde{\mathcal{F}}(q_{0})=\tilde{\mathcal{F}}(-q_{0})=-\frac{i\alpha}{16\pi q_{0}^{3}},\qquad\alpha:=\langle u,W\rangle,
\end{equation}
and the prediction for the shelf height is 
\begin{equation}
    \tilde{q}^{-}\sim-\frac{\alpha}{4q_{0}^{3}}.
\end{equation}
The parameter $\alpha$ 
defined in \eqref{e:alpha} 
depends on the form of the perturbation $F$ through $W$. 
Note that $\alpha$ should be computed \textit{after} incorporating the time evolution of the parameters determined from the orthogonality condition \eqref{e:orth} into $W$. Importantly, no shelf is expected to develop for any perturbation such that 
$\alpha(T)\equiv 0$.

\section{Applications}

\begin{figure}[!htb]
\centering
    \includegraphics[width=\textwidth]{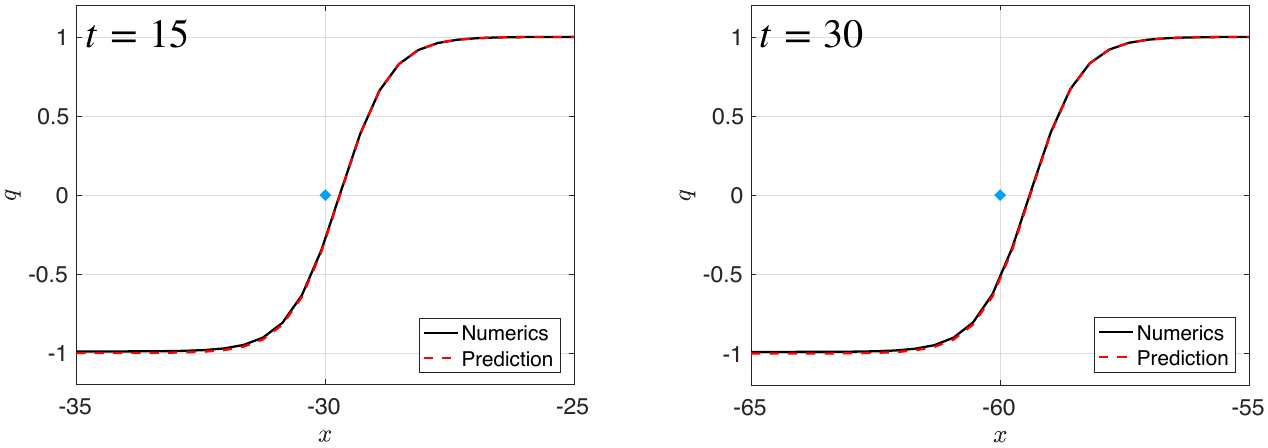}
    \caption{Comparison of the predicted (dashed red line) and numerical (solid black line) solutions for a kink under the influence of the diffusion perturbation $F[q]=q_{xx}$ with $\varepsilon=0.03$ and fixed times $t=15$ (left) and $t=30$ (right). The initial parameters are $q_{0}(0)=1$ and $x_{0}(0)=0$. The blue marker indicates the location of the kink center in the absence of the perturbation.}
    \label{f:diff_snapshots}
\end{figure}

\begin{figure}[!htb]
\centering
    \includegraphics[width=\textwidth]{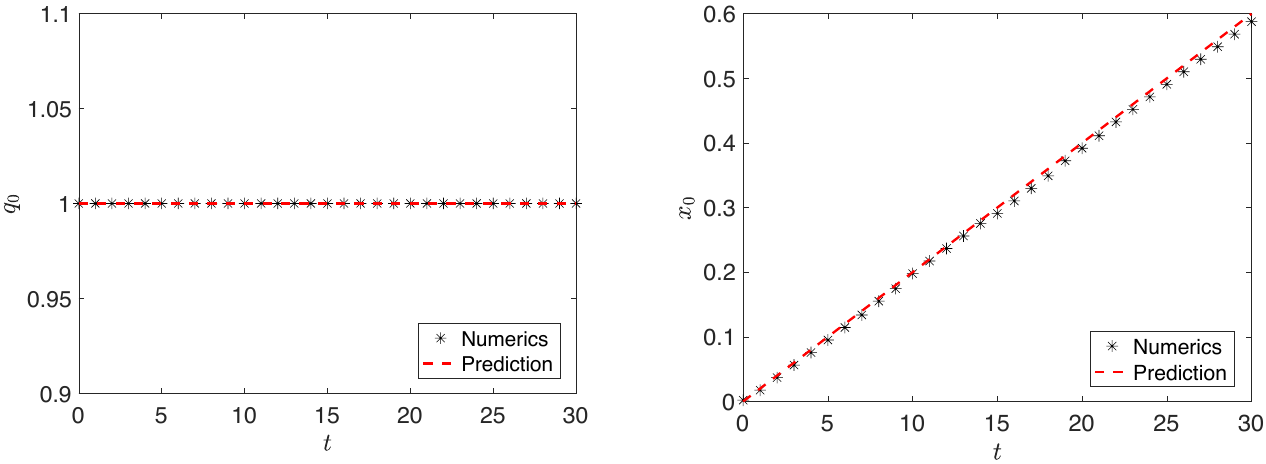}
    \caption{Predicted evolution of the background amplitude $q_{0}$ (left) and kink center $x_{0}$ (right) with time (dashed red line) under the influence of the diffusion perturbation compared with numerical measurements (black stars). The parameters are the same as in Fig.~\ref{f:diff_snapshots}.}
    \label{f:diff_parameters}
\end{figure}

In this section, we apply our results to certain physically relevant perturbations and show that the predictions are consistent with direct numerical simulations. First, consider a diffusive perturbation of the form
\begin{equation}
    F[q]=q_{xx}.
\end{equation}
In this case, the orthogonality condition \eqref{e:orth} reads
\begin{equation}
    q_{0T}\int_{-\infty}^{\infty}e^{\xi}\sech\xi\tanh\xi \,d\xi+q_{0T}+\frac{4}{3}q_{0}^{3}-2q_{0}^2x_{0T}=0.
\end{equation}
The remaining integral above is divergent, so it must be the case that $q_{0T}=0$, i.e. the background amplitude does not evolve on the slow time scale. Note that this is consistent with \eqref{e:q0T}. With this, the remaining terms give a formula for the evolution of the center
\begin{equation}
    x_{0T}=\frac{2}{3}q_{0}.
\end{equation}
Thus, the perturbed kink can be approximated near its center as
\begin{equation}
\label{e:dif_approx}
    q(x,t)\approx q_{0}(0)\tanh\left[q_{0}(0)\left(x+2q_{0}(0)^2t-x_{0}(0)-\frac{2}{3}q_{0}(0)T\right)\right].
\end{equation}
In Fig.~\ref{f:diff_snapshots}, comparisons between the predicted form of the kink \eqref{e:dif_approx} and a numerical simulation at fixed instants in time are shown. The numerically tracked kink parameters are plotted alongside the corresponding predictions in Fig.~\ref{f:diff_parameters}, showing that the predictions hold up well even at times of order $1/\varepsilon$. 


\begin{figure}[!htb]
\centering
    \includegraphics[width=\textwidth]{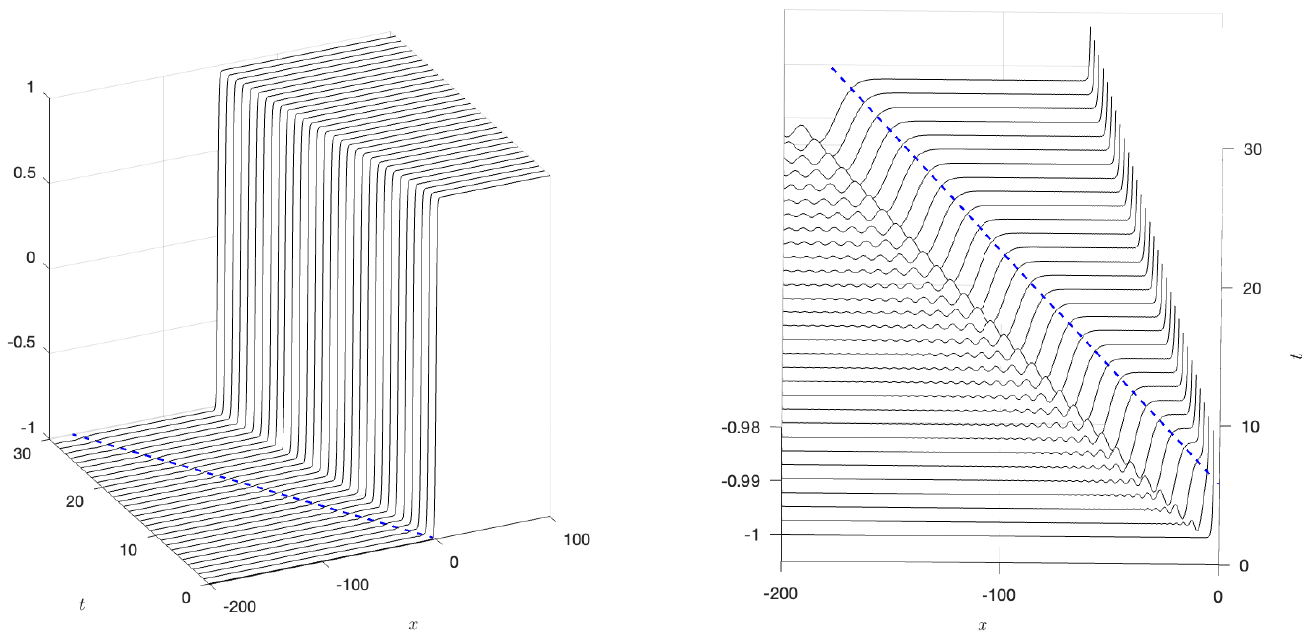}
    \caption{A spacetime plot of the evolution of a kink under the influence of the diffusion perturbation. The right panel is a zoomed in view showing the propagation of the shelf in front of the kink. The blue dashed line denotes the predicted boundary of the shelf region. The parameters are the same as in Fig.~\ref{f:diff_snapshots}.}
    \label{f:diff_spacetime}
\end{figure}

\begin{figure}[!htb]
\centering
    \includegraphics[width=\textwidth]{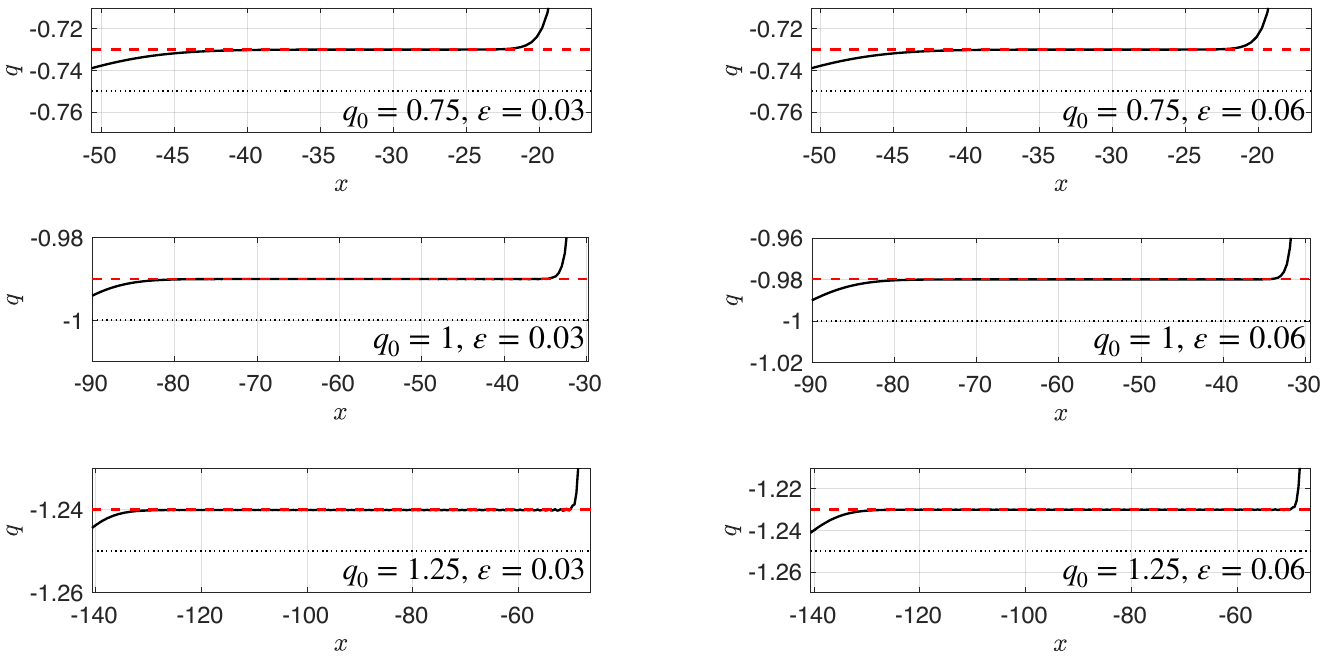}
    \caption{Snapshots at $t=15$ of the numerical simulation (solid black line) in the shelf region alongside the asymptotic prediction $-q_{0}+\varepsilon/3$ (dashed red line) for varying $q_{0}$ and $\varepsilon$.}
    \label{f:diff_shelf}
\end{figure}

Furthermore, we find that $\alpha=\langle u,W\rangle=-4q_{0}^{3}/3$, from which the prediction for the height of the shelf on the left side of the kink is $\tilde{q}^{-}\sim1/3$. That is, in the shelf region we have
\begin{equation}
    q(x,t)\sim -q_{0}+\frac{\varepsilon}{3}.
\end{equation}
Figure~\ref{f:diff_spacetime} shows a spacetime plot of the evolution of the perturbed kink, in which the propagation of the shelf can be seen. Interestingly, the shelf height is independent of the background amplitude. This is corroborated by numerical evidence as shown in Fig.~\ref{f:diff_shelf}.

\begin{figure}[!htb]
\centering
    \includegraphics[width=\textwidth]{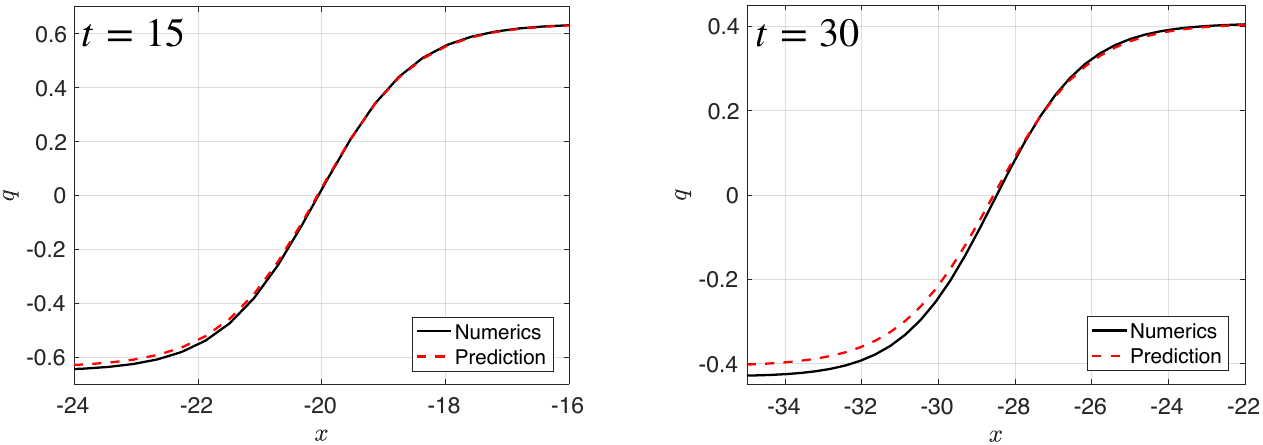}
    \caption{Comparison of the predicted (dashed red line) and numerical (solid black line) solutions for a kink under the influence of the damping perturbation $F[q]=-q$ with $\varepsilon=0.03$ and fixed times $t=15$ (left) and $t=30$ (right). The initial parameters are $q_{0}(0)=1$ and $x_{0}(0)=0$. Note that the development of the shelf is visible on the left side of the kink.}
    \label{f:damp_snapshots}
\end{figure}

\begin{figure}[!htb]
\centering
    \includegraphics[width=\textwidth]{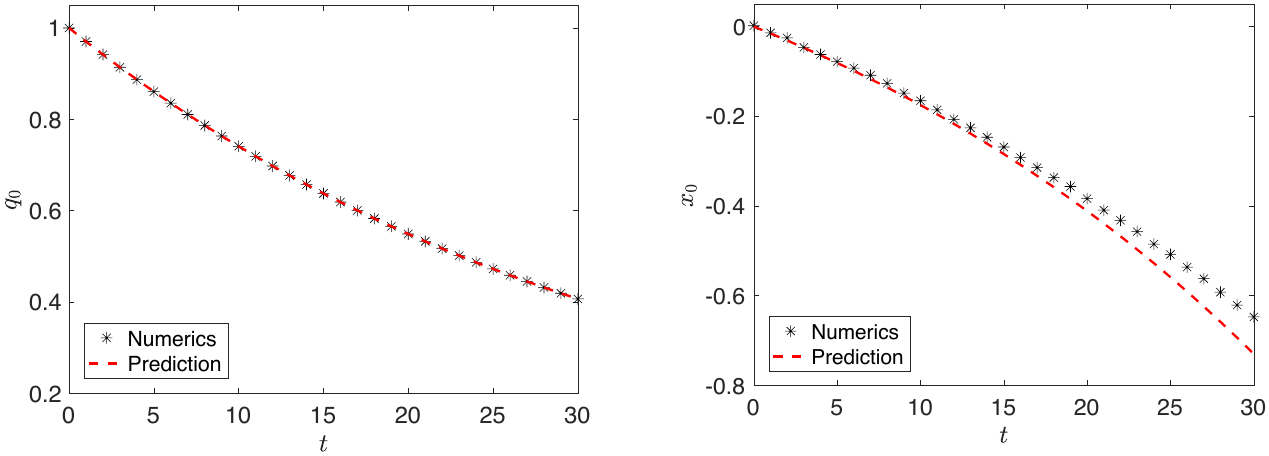}
    \caption{Predicted evolution of the background amplitude $q_{0}$ (left) and kink center $x_{0}$ (right) with time (dashed red line) under the influence of the damping perturbation compared with numerical measurements (black stars). The parameters are the same as in Fig.~\ref{f:damp_snapshots}.}
    \label{f:damp_parameters}
\end{figure}

Next, consider the linear damping/loss perturbation 
\begin{equation}
    F[q]=-q.
\end{equation}
The orthogonality condition \eqref{e:orth} is
\begin{equation}
    (q_{0}+q_{0T})\int_{-\infty}^{\infty}e^{\xi}\sech\xi\tanh\xi \,d\xi+q_{0T}-2q_{0}^{2}x_{0T}=0.
\end{equation}
To remove the divergent integral, the background amplitude must evolve according to
\begin{equation}
    q_{0T}=-q_{0}.
\end{equation}
The remaining terms show that the kink center satisfies
\begin{equation}
    x_{0T}=-\frac{1}{2q_{0}}.
\end{equation}
Therefore, the parameters are given explicitly by
\begin{equation}
    q_{0}(T)=q_{0}(0)e^{-T},\qquad x_{0}(T)=x_{0}(0)-\frac{1}{2q_{0}(0)}\big(e^{T}-1\big),
\end{equation}
and the kink is approximated near its center by
\begin{equation}
\label{e:damp_approx}
    q(x,t)\approx q_{0}(0)e^{-T}\tanh\left[q_{0}(0)e^{-T}\left(x+q_{0}(0)^{2}\frac{1-e^{-2T}}{\varepsilon}-x_{0}(0)+\frac{1}{2q_{0}(0)}\big(e^{T}-1\big)\right)\right].
\end{equation}
Comparisons between \eqref{e:damp_approx} and a numerical simulation at fixed times are displayed in Fig.~\ref{f:damp_snapshots}, and the numerically measured evolution of the parameters are shown in Fig.~\ref{f:damp_parameters}. 

\begin{figure}[!htb]
\centering
    \includegraphics[width=\textwidth]{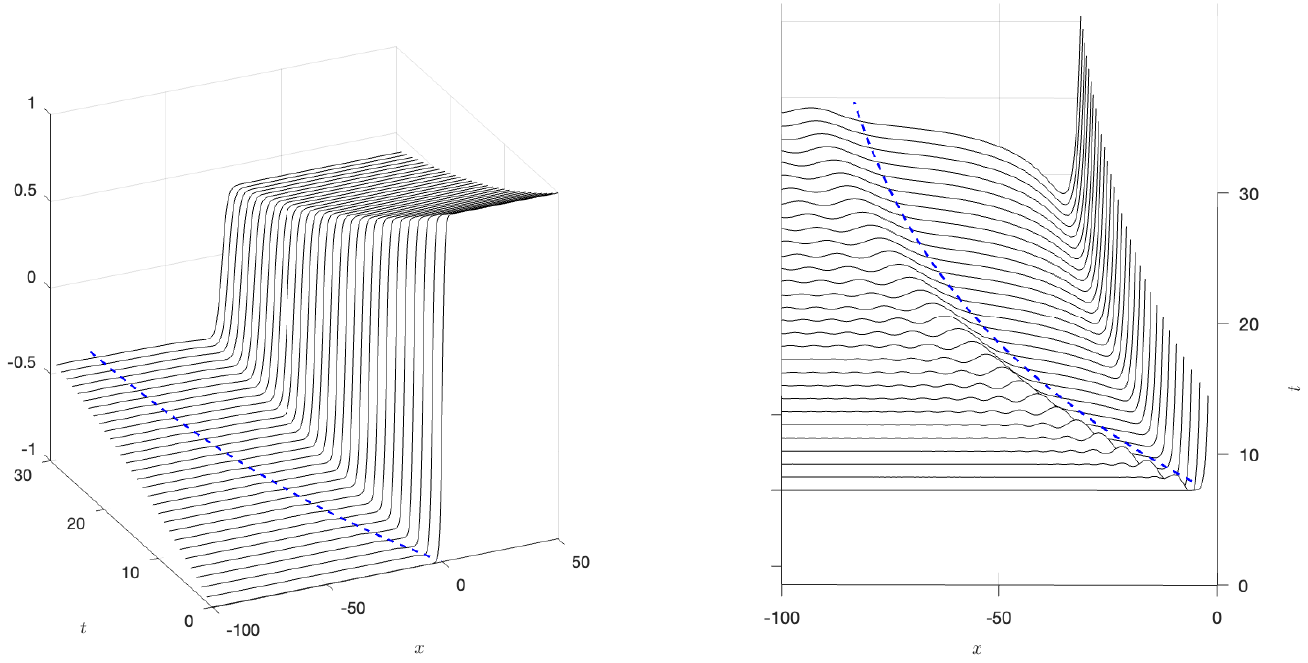}
    \caption{A spacetime plot of the evolution of a kink under the influence of the damping perturbation. The right panel is a zoomed in view (with the background evolution artificially removed) showing the propagation of the shelf in front of the kink. The blue dashed line denotes the predicted boundary of the shelf region. The parameters are the same as in Fig.~\ref{f:damp_snapshots}.}
    \label{f:damp_spacetime}
\end{figure}

The shelf parameter in this case is found to be $\alpha=q_{0}$ so that the shelf height is $\tilde{q}^{-}\sim-1/(4q_{0}^{2})$, i.e. in the shelf region
\begin{equation}
    q(x,t)\sim -q_{0}-\frac{\varepsilon}{4q_{0}^{2}}.
\end{equation}
Note that in this case the left edge of the shelf is given by 
$-6\int_{0}^{t}q_{0}(\varepsilon s)^2ds$. A spacetime plot showing the slow decrease in the background amplitude and the development the shelf is shown in Fig.~\ref{f:damp_spacetime}.

In closing, we mention again that while the development of a prominent shelf in front of the kink is one of the key results of this analysis, the shelf is not generic. For any perturbation such that $\alpha=0$, the first order correction integral is asympotically negligible throughout the spatial domain. Using \eqref{e:q0T} for convenience, the parameter $\alpha$ can be written as
\begin{equation}
    \alpha=\int_{-\infty}^{\infty}\tanh\xi\Big\{F[u(\xi)]-F[q_{0}]\tanh\xi\Big\}d\xi-F[q_{0}].
\end{equation}
The above equation indicates that all perturbations such that 
$F[q_{0}]=0$ and $F[u(\xi)]$ is even are included in the class of perturbations for which no shelf develops. This property is held by such physically relevant effects as higher order dispersion of the form $F[q]=\partial_{x}^{2n+1}q$ and higher order nonlinearity of the form $F[q]=q^{2n}q_{x}$. It is also worth noting that if one defines the renormalized energy to be $E=(1/2)\int_{-\infty}^{\infty}(q_{0}^2-q(x,t)^2)dx$, then the following simple relationship between the evolution of the energy, the evolution of the background, and the shelf holds:
\begin{equation}
    E_{T}=q_{0T}-\alpha.
\end{equation}
The physical intuition that can be extracted from this equation is that any change to the renormalized energy of a perturbed kink must arise from a change to the background amplitude and/or from the development of a shelf. Moreover, for any conservative perturbation, these effects must exactly balance each other. On the other hand, for a dissipative perturbation that does not allow for evolution of the background (as is the case for the diffusion perturbation discussed above), the energy loss should be associated with the emergence of  a shelf.

\section{Conclusion}

In this paper we have developed an integrable perturbation theory for the kink solution of the defocusing mKdV equation, based on the squared eigenfunction expansion associated with the underlying Zakharov–Shabat scattering problem. Starting from the completeness relation for the squared eigenfunctions on the kink background — with careful treatment of the singularities of the scattering data at the branch points of the continuous spectrum — we established the orthogonality conditions, and obtained explicit slow-time evolution equations for the perturbed kink parameters at leading order. The resulting framework provides a rigorous foundation for analyzing the response of the mKdV kink to a broad class of physically relevant perturbations, including dissipative, forcing, and higher-order dispersive terms.

Several natural sequels of this work suggest themselves. The most immediate is the application of the same squared-eigenfunction framework to the dark soliton of the defocusing mKdV equation, which, like the kink, lives on a nonzero background but has a different topological character. The main difference compared to the kink (and to the dark soliton of the defocusing NLS equation in the same hierarchy as mKdV) is the fact that the spectrum of the mKdV dark soliton is characterized by 4 symmetric discrete eigenvalues (2 in each half-plane of the spectral parameter). Consequently, the set of discrete squared eigenfunctions entering the closure relation is correspondingly doubled. 
Although the presence of additional discrete modes increases the algebraic complexity of the calculations, we expect the analysis to proceed along essentially the same lines as in the present work. In particular, it should closely parallel the perturbation theory recently developed for the dark soliton of the defocusing NLS equation in \cite{OPY26}, with the extra modes requiring careful bookkeeping, but introducing no fundamental difficulties.
The treatment of the branch-point singularities of the scattering data, which was the critical technical ingredient here, is expected to carry over in the same form as for the kink.

A further direction concerns the extension of the present perturbation theory to the recently introduced integrable fractional mKdV equation. Ablowitz, Been, and Carr \cite{ABC2022} showed that the IST can be extended to fractional nonlinear evolution equations characterized by anomalous dispersion, using completeness of suitable eigenfunctions of the associated linear scattering problem, and derived in particular an integrable fractional mKdV equation whose solitons exhibit anomalous dispersion. Since the squared eigenfunction completeness relation is the key structural ingredient both in the construction of those fractional hierarchies and in the perturbation theory developed here, it is natural to ask whether the present framework can be adapted to study perturbations of kink solutions in the fractional setting, where the interplay between anomalous dispersion and topological structure is largely unexplored.

\section*{Acknowledgments}
BP and NJO gratefully acknowledge partial support for this work from the NSF, under grant DMS-2406626. BP also acknowledges the Fulbright Foundation in Greece and the Fulbright program, and the Mathematics Department of the University of Ioannina, Greece, for the kind hospitality during the completion of this work. 

\section*{Conflict of interest}
The authors declare that they have no conflict of interest.

\end{document}